\documentclass[10pt,letterpaper]{article}
\usepackage{opex3}
\usepackage{hangcaption}
\usepackage{array}
\usepackage{indentfirst} 
\usepackage{cite}

\begin{document}




\title{Random phase-free \\ 
computer-generated hologram}

\author{
Tomoyoshi Shimobaba and  Tomoyoshi Ito
}

\address{Graduate School of Engineering, Chiba University, 1-33 Yayoi-cho, Inage-ku, Chiba 263-8522, Japan
}
\email{$^*$shimobaba@faculty.chiba-u.jp}

\begin{abstract} 
Addition of random phase to the object light is required in computer-generated holograms (CGHs) to widely diffuse the object light and to avoid its concentration on the CGH; however, this addition causes considerable speckle noise in the reconstructed image. For improving the speckle noise problem, techniques such as iterative phase retrieval algorithms and multi-random phase method are used; however, they are time consuming and are of limited effectiveness. Herein, we present a simple and computationally inexpensive method that drastically improves the image quality and reduces the speckle noise by multiplying the object light with the virtual convergence light. Feasibility of the proposed method is shown using simulations and optical reconstructions; moreover, we apply it to lens-less zoom-able holographic projection. The proposed method is useful for the speckle problems in holographic applications.
\end{abstract}


\


\section{Introduction}
\noindent Computer-generated hologram (CGH) \cite{goodman2005introduction} includes a unique capability for recording and reconstructing the desirable amplitude and phase of object light, leading to a wide use of this technique in optical applications such as three-dimensional (3D) displays \cite{sasaki2014large}, projections \cite{buckley2011holographic}, multi-spot generation \cite{Ogura:14}, diffractive optical elements \cite{kress2000digital}, and encryption \cite{alfalou2009optical}.
Random phase addition to the object light has been required to diffuse the object light and to avoid the concentration of the object light on the CGH since the initial development of CGHs \cite{Lohmann:67} in the 1960s; 
however, this causes considerable speckle noise in the reconstructed image.
Consequently, iterative phase retrieval algorithms \cite{doi:10.1117/1.2148980}, multi-random phase method \cite{Amako:95} and pixel separation methods \cite{Takaki:11, Makowski:13, Mori:14} are used for improving the speckle noise problem; unfortunately, these are time consuming and of limited effectiveness. 
 
In this paper, we present a simple and computationally inexpensive method called ``random phase-free computer-generated hologram'' that drastically improves the image quality and reduces the speckle noise by multiplying the object light with the virtual convergence light.
Furthermore, we show the feasibility of the proposed method using simulations and optical reconstructions, and apply to lens-less zoom-able holographic projection \cite{shimobaba2013lensless,ducin}.
The proposed method is useful for the speckle problems in a wide-range of  holographic applications.
Error diffusion methods \cite{ed1,ed2,ed3} and down-sampling method \cite{down} are known as existing random phase-free methods; unfortunately, these methods cannot widely diffuse the object lights, so that it is difficult to apply these methods to lens-less zoom-able holographic projection \cite{shimobaba2013lensless,ducin}.
Whereas, the proposed method can use in the application.

\section{Proposed method}
\begin{figure}
\centerline{\includegraphics[width=12cm]{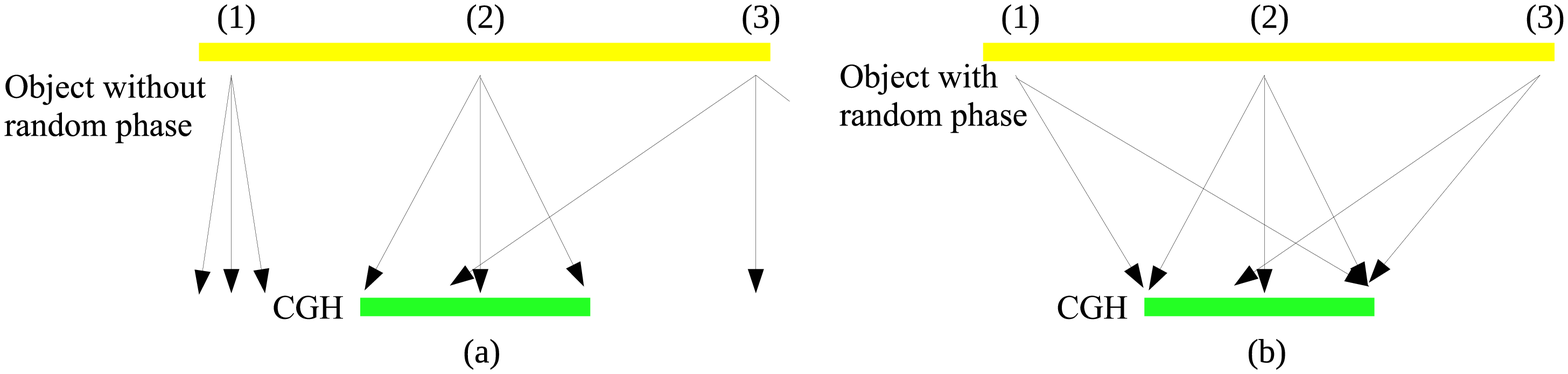}}
\caption{Illustration of the intuitive reason for the random phase addition. (a) without random phase (b) with random phase.}
\label{fig:random}
\end{figure}

\noindent
To begin with, we discuss an intuitive reason for the use of random phase. 
A schematic illustration of the CGH calculation with and without the random phase is presented in Fig. \ref{fig:random}, respectively.
Consider a two-dimensional image, consisting of parts with low spatial frequencies ((1) and (2) in Fig. \ref{fig:random}) and high spatial frequencies ((3) in Fig. \ref{fig:random}).
In addition, most images in general contain many low frequency components, and thus cannot spread light widely.
In the case without random phase, shown in Fig. \ref{fig:random}(a),  the  low frequency object parts cannot spread light widely over the CGH because the spread angle is proportional to $\sin^{-1}(\lambda \nu)$ where $\lambda$ and $\nu$ are the wavelength and the spatial frequency, respectively.
Therefore, the object information of parts (2) and (3) can be recorded on the CGH, whereas that of part (1) cannot be recorded. 
Conversely, as shown in Fig. \ref{fig:random}(b), random phase acts as an equivalent of a physical diffuser so that light from all parts of the object can spread over the CGH owing to high spatial frequency provided by the random phase.
Therefore, the information of the object light can be recorded on the CGH; however, considerable speckle noise is present in the reconstructed image  because of random interference by the diffused light. 

\begin{figure}
\centerline{\includegraphics[width=9cm]{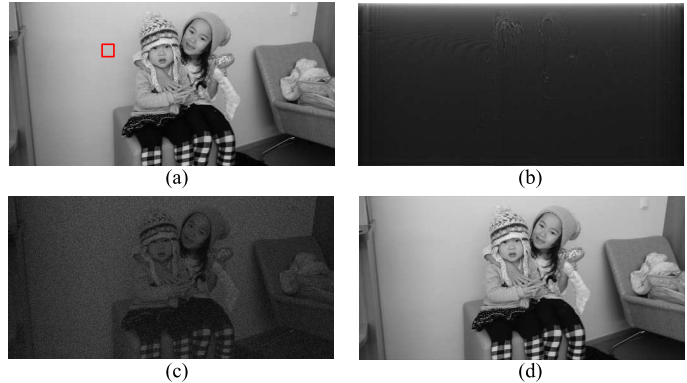}}
\caption{Original image (a). Numerically reconstructed image from 4K resolution CGHs without the random phase (b). Numerical reconstructed image with the random phase (c). Numerical reconstructed image with the proposed method (d). The red box indicates the region of interest for measuring speckle contrast.}
\label{fig:random_image}
\end{figure}

Next, we show the effects of the absence and the presence of the random phase in Fig. \ref{fig:random_image} as obtained by simulations.
Here, we assume an amplitude-modulated spatial light modulator (SLM) with the resolution of $4,096 \times 2,048$ pixels (henceforth referred to as 4K resolution) and pixel pitch of $8.5\mu$m. 
The wavelength of the reference light is 532 nm, and the distance between the object and CGH is 0.8 m.
Figure \ref{fig:random_image}(a) represents the original image and 
Fig. \ref{fig:random_image}(b) is the reconstructed image from the CGH without the random phase.
From Fig. 2(b), it is clear that only a part of the original image is reconstructed.
Moreover, Fig. \ref{fig:random_image}(c) is the reconstructed image from the CGH with random phase.
Here, we can observe the whole image; however, it is contaminated with speckle noise.

In order to reduce this speckle noise, the multi-random phase method \cite{Amako:95} was proposed.
This method reduces the speckle noise by averaging the reconstructed images with different speckle noise from multiple CGHs generated by different random phases with a high-speed displaying device.
The speckle reduction is proportional to $\sqrt{N}$ where $N$ is the number of CGHs.
For example, to improve by a factor of ten, one hundred CGHs would be required.

Recently, an efficient speckle reduction method known as the pixel separation was proposed \cite{Takaki:11,Makowski:13, Mori:14}.   
 It is known that adjacent pixels of the reconstructed images interfere with each other because the point spread function of the reconstructed pixel has side lobes; consequently, such interferences cause speckle noise.
As a result, in the pixel separation method, to avoid the unnecessary interference of the side lobes, the pixels in the object are separated from each other, and CGHs are generated from the sparse objects.
The speckle-reduced image is reconstructed by switching the CGHs generated from the sparse objects at high-speed.

Currently, the major speckle reduction methods are either based on an iterative algorithm \cite{doi:10.1117/1.2148980}, where the CGH pattern is optimized by iterating the calculations using known information, or the multiple-CGH method; however they are time-consuming and their effectiveness is limited. 
Additionally, the use of the multiple-CGH method requires special display devices with high-speed refresh rates.

\begin{figure}
\centerline{\includegraphics[width=6cm]{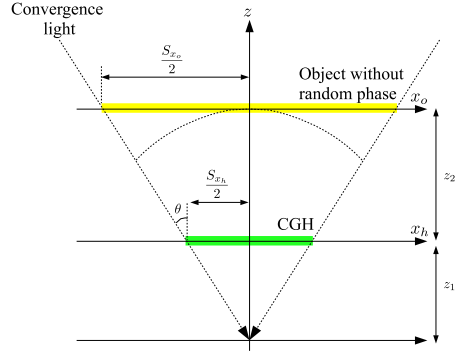}}
\caption{Random phase-free CGH using convergence light.}
\label{fig:random_free}
\end{figure}

Keeping the above limitations in mind, we propose a random phase-free CGH method that does not require time-consuming processing or special display devices.
Figure \ref{fig:random_free} shows the calculation setup for the random phase-free CGH.
The complex amplitude of the object and the CGH planes are $u_o(x_o, y_o)$ and $u_h(x_h, y_h)$. 
Instead of using the random phase, we multiply $u_o(x_o, y_o)$ by the virtual convergence light given by $w(x_o, y_o)$.
To avoid the aliasing error, the incident angle $\theta$ of the convergence light must satisfy $\theta=\sin^{-1}(\lambda/2 p_h)$ where $\lambda$ is the wavelength and $p_h$ is the sampling rate on the CGH, respectively.
The distance from the focus point of the convergence light is denoted by $z_1$, and is set to the distance at which the CGH just fits to the cone of the convergence light.
The recording distance (and the reconstruction distance) are denoted by $z_2$.
Subsequently, we calculate the complex amplitude on the CGH using $u_h(x_h, y_h) = {\rm Prop_{z_2}}\{u_o(x_o, y_o) w(x_o, y_o)\}$ where $ {\rm Prop_{z_2}\{\cdot \}}$ denotes light propagation, such as a diffraction calculation \cite{goodman2005introduction}, at the propagation distance $z_2$.
Current major displaying devices are amplitude-modulated and phase-modulated SLMs; therefore the real part of the $u_h(x_h, y_h)$ complex amplitude is relevant for amplitude-modulated SLMs or the argument of $u_h(x_h, y_h)$ for the phase-modulated SLMs. 

Now, we describe how to determine the virtual convergence light $w(x_o, y_o)$.
When the horizontal and vertical sizes of the object and the CGH are the same, i.e., $S_{x_o} \times S_{x_o}$ and $S_{x_h} \times S_{x_h}$, respectively, $w(x_o, y_o)$ is expressed as a virtual convergence light $w(x_o, y_o)=\exp(-i \pi(x_o^2+y_o^2)/\lambda f)$ where $f=z_1+z_2$ is the focal length.
Using a simple geometric relation as shown in Fig.\ref{fig:random_free}, we can derive $S_{x_h}/2 : S_{x_o}/2 = z_1: f$, hence $f=z_2/(1-S_{x_h}/ S_{x_o})$.

Conversely, when the horizontal and vertical sizes of the object and CGH are not the same, $S_{x_o} \times S_{y_o}$ and $S_{x_h} \times S_{y_h}$ respectively, $w(x_o, y_o)$ is expressed as an anamorphic convergence  light $w(x_o, y_o)=\exp(-i \pi(x_o^2/\lambda f_x+y_o^2/\lambda f_y))$ where $f_x$ and $f_y$ denote the focal lengths where $f_x=z_2/(1-S_{x_h}/ S_{x_o})$ and $f_y=z_2/(1-S_{y_h}/ S_{y_o})$.
To avoid an overlap between the object and the direct light, the original object must be shifted from the optical axis by the distance $(o_x,o_y)$.
This is also true for the methods used to obtain Fig.\ref{fig:random_image}.
Owing to the addition of the shift amount, the focal lengths of the convergence light are $f_x=z_2/(1-S_{x_h}/ (S_{x_o} + 2 o_x))$ and $f_y=z_2/(1-S_{y_h}/ (S_{y_o} + 2 o_y))$, respectively.

\section{Results}
\noindent Despite its simplicity, our proposed method is quite powerful.
Figure \ref{fig:random_image}(d) shows an image that was numerically reconstructed using the proposed method while retaining the same calculation conditions used in Fig.\ref{fig:random_image}. 

As we can see, the image quality is dramatically higher in Fig.\ref{fig:random_image}(d) than in Figs.\ref{fig:random_image}(b) and (c).
The peak signal-to-noise ratios (PSNRs) between the original image of Fig.\ref{fig:random_image}(a) and the reconstructed images of Figs.\ref{fig:random_image}(b), \ref{fig:random_image} (c) and \ref{fig:random_image}(d) were 5.6 dB, 8.8dB and 37.4dB, respectively.
Generally speaking, human eyes cannot distinguish between two images with PSNR over 30 dB.
Next, we estimate the speckle reduction effect of the proposed method using the speckle contrast  \cite{Amako:95}  $C=\sigma/{\rm<}I{\rm>}$ where $\sigma$ and ${\rm<}I{\rm>}$ are the standard deviation and the average for the region of interest (ROI).
Although a speckle contrast below 5\% may be acceptable in many cases,  for consumer products, a speckle contrast below 1\% is usually desirable \cite{Manni:12}.
The ROI with an area of $100 \times 100$ pixels is denoted by the red box in Fig. \ref{fig:random_image} (a).
The speckle contrasts are about 24 \%, 50 \% and 1 \% for Figs. \ref{fig:random_image} (b),  \ref{fig:random_image}(c) and \ref{fig:random_image}(d), respectively.
Thus, the proposed method drastically improves both the PSNR and the speckle contrast.

%


\begin{figure}
\centerline{\includegraphics[width=10cm]{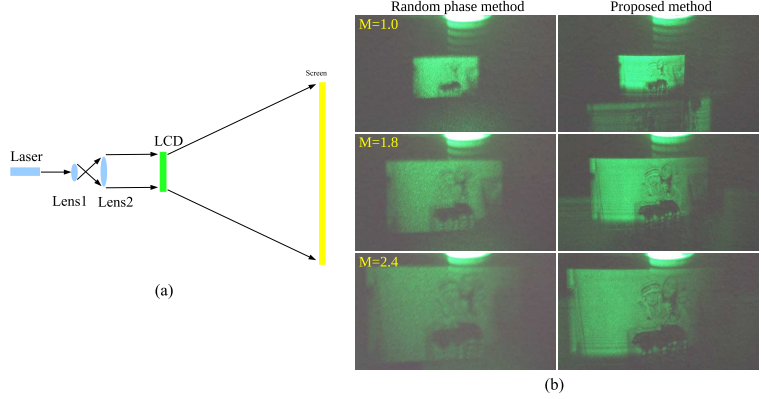}}
\caption{Optical system (a). Optical reconstructions by the random phase method and the proposed method (b). The magnification changes are 1.0, 1.8 and 2.4.}
\label{fig:opt_image}
\end{figure}

Although the proposed method can be used in wide-ranging holography, we focus on lens-less zoom-able holographic projection \cite{shimobaba2013lensless, ducin} in order to show the feasibility of the proposed method by optical experiments.
Unlike the other projection methods, holographic projection intrinsically does not need any lens including the zoom lens.
Since it only requires an SLM and a light source and does not require any optical components, the lens-less zoom-able holographic system is extremely simple, and hence is a promising technique for use in ultra-small projectors.
We have previously developed a similar system where a numerical scaled Fresnel diffraction method\cite{shimobaba:2013} that calculates the diffraction at different sampling rates on an object and on the CGH is used instead of the zoom lens. 

To zoom a projected image using the scaled Fresnel diffraction, a larger sampling rate $p_o$ must be used on an object, than on a CGH.
In our previous work \cite{shimobaba2013lensless}, we multiplied the object function by the random phase to widely diffuse the object light over the CGH.
As mentioned earlier, random phase is one of the main causes of the speckle noise; therefore, we used the time integration of multiple CGH with different speckle noises to reduce it.
Unfortunately, this procedure was time consuming and of limited effectiveness.
We now replace the random phase with the proposed method in our lens-less zoom-able holographic projection.

We calculate the complex amplitude on the CGH by $u_h(x_h, y_h) = {\rm Prop^{p_o,p_h}_{z_2}}\{u_o(x_o, y_o) w(x_o, y_o)\}$ where the operator $ {\rm Prop ^{p_o,p_h}_{z_2}\{\cdot \}}$ denotes scaled diffraction calculations at different sampling rates.
The optical system is shown in Fig. \ref{fig:opt_image}(a).
We used an amplitude-modulated liquid crystal display (L3C07U made by EPSON) with the resolution of $1,920 \times 1,080$ pixels and the pixel pitch of 8.5$\mu$m.
To generate the amplitude-modulated CGHs, we take the real part of the complex amplitude on the CGH.
The light source is a semiconductor laser with a wavelength of 532 nm and the power of 120 mW.
The collimator consists of two lenses.
The calculation conditions are  $z_2=0.8$m, $p_h=8.5\mu$m and $p_o=M p_h$ where $M$ is the magnification of the reconstructed image. 
 We captured the reconstructed images on the screen using a consumer digital camera.


Figure \ref{fig:opt_image}(b) shows the optical reconstructions from CGHs by the random phase and proposed methods.
The magnification $M$ values were 1.0, 1.8 and 2.4, respectively.
Furthermore, we show some movies recorded using the proposed method  (\textcolor{blue}{movie 1}) and random phase method (\textcolor{blue}{movie 2}) with the magnification of 2.4.
Both reconstructions can be zoomed; however, the optical reconstruction using the proposed method is of higher image quality than that using the random phase method, with lower speckle noise and a sharper image.
The results using the proposed method contain speckle noise; however, the speckle noise is not caused by the proposed method, but is rather caused by the rough surface of the screen.
This is a common problem in laser projectors and can be ameliorated by vibrating the screen or by using a low speckle noise laser source, for example.
All of the calculations in this paper were done by our numerical library for wave optics \cite{Shimobaba20121124}.

\section{Conclusion}
\noindent  To summarize, we proposed a random phase-free CGH and showed its feasibility using simulation and optical experiments. 
The proposed method drastically improves the PSNR and the speckle contrast. 
In the future, we will investigate the feasibility of the proposed method for holographic 3D display, encryption, and multiple-spot generation.

\section*{Acknowledgments}
\noindent This work is supported by Japan Society for the Promotion of Science (JSPS) KAKENHI (Grant-in-Aid for Scientific Research (C) 25330125) 2013, and KAKENHI (Grant-in-Aid for Scientific Research (A) 25240015) 2013.
\end{document}